\pdfoutput=1
\documentclass{naturRevtex}

\usepackage{units,color,import}
\usepackage{graphicx}
\usepackage{dcolumn}
\usepackage{bm}
\usepackage{graphicx, datetime, bbold,multirow}
\usepackage{amsmath,MnSymbol,mathtools,amsfonts}
\usepackage{psfrag,comment}
\usepackage{caption,subcaption}
\usepackage[percent]{overpic}

\usepackage{CustomMath}
\usepackage{CustomFigFont}

\date{\today}
\begin{document}

\title{Entanglement-enhanced optical gyroscope}

\author{Matthias Fink$^{1,2,*}$, Fabian Steinlechner$^{1,2,3}$, Johannes Handsteiner$^{1,2}$, Jonathan P. Dowling$^{4,5,6,7}$, Thomas Scheidl$^{1,2}$ \& Rupert Ursin$^{1,2,*}$}

\DeclareGraphicsExtensions{.pdf,.jpg,.eps} 

\maketitle

\begin{affiliations}
 \item Institute for Quantum Optics and Quantum Information - Vienna (IQOQI), Austrian Academy of Sciences, Vienna, Austria
 \item Vienna Center for Quantum Science and Technology (VCQ), Vienna, Austria
 \item Current address: Fraunhofer Institute for Applied Optics and Precision Engineering IOF, Jena, Germany
  \item Hearne Institute for Theoretical Physics, Department of Physics and Astronomy, Louisiana State University, Baton Rouge, Louisiana 70803, USA
 \item CAS-Alibaba Quantum Computing Laboratory, USTC, Shanghai 201315, China
 \item NYU-ECNU Institute of Physics at NYU Shanghai, Shanghai 200062, China
 \item National Institute of Information and Communications Technology, Tokyo 184-8795, Japan
\end{affiliations}

\begin{abstract}

Fiber optic gyroscopes (FOG) based on the Sagnac effect are a valuable tool in sensing and navigation and enable accurate measurements in applications ranging from spacecraft\cite{wang:2012} and aircraft\cite{sanders:2002} to self-driving vehicles such as autonomous cars\cite{kajioka:1996}. As with any classical optical sensors, the ultimate performance of these devices is bounded by the standard quantum limit (SQL). Quantum-enhanced interferometry allows us to overcome this limit using non-classical states of light. Here, we report on an entangled-photon gyroscope that uses path-entangled NOON-states ($N=2$) to provide phase supersensitivity beyond the standard-quantum-limit.



\end{abstract}
\vspace{\baselineskip}

\section*{Introduction}

Among the many applications of optical interferometry, optical gyroscopes based on the Sagnac effect are an invaluable tool in sensing and navigation. First observed by Georges Sagnac in 1913 \cite{sagnac:1913:exp,sagnac:2008:exp} --- in an attempt to observe the "relative circular motion of the luminiferous ether within the closed optical path"\cite{sagnac:1913:proof,sagnac:2008:proof} --- the Sagnac effect refers to the relative phase $\phi_S(\Omega)$ experienced by counter-propagating light waves in a rotating interferometer. To this day, this experiment, together with that of Michelson and Morley\cite{michelson:1887}, is considered one of the fundamental experimental tests of the theory of relativity\cite{Logunov:1988}. While the relativistic correction of the effect is still under discussion\cite{malykin:2000,malykin:2002,ghosal:2003,gift:2018}, a mass product has evolved from its application\cite{kajioka:1996}. The effect allows us to determine the absolute rotation $\Omega$ with respect to inertial space \cite{malykin:2000} and has since found application in navigation systems for spacecraft\cite{wang:2012} and aircraft\cite{sanders:2002} as well as self-driving vehicles such as autonomous cars \cite{kajioka:1996}.

The precision of an optical gyroscope is determined by the phase response $\frac{\partial\phi}{\partial\Omega}$ as well as the minimum phase resolution $\Delta \phi$. The phase response $\frac{\partial\phi}{\partial\Omega} = S_{T} \propto \frac{A_{\closedpath}}{\lambda}$, or Sagnac scale factor $S_{T}$, is proportional to the area $A_{\closedpath}$ enclosed by the counter propagating waves and inverse proportionality to the wavelength $\lambda$ of the interfering wave. In commercial fiber optic gyroscopes (FOG), the phase response is amplified by increasing the effective area enclosed by the optical paths by using an optical fiber-coil. Another strategy to improve the precision is to use shorter wavelengths. Therefore, the Sagnac effect was examined with X-rays\cite{vysotskii:1994} as well as with de Broglie waves such as electrons\cite{hasselbach:1993}, neutrons\cite{werner:1979}, and atoms\cite{riehle:1991,lenef:1997,gustavson:1997}. The generation and guiding of such waves, however, is rather difficult in comparison to optical electromagnetic waves, and the area enclosed by such gyroscopes is rather small compared to the FOG.

Furthermore, approaching the ultimate limits of sensitivity, the minimum phase that can be measured using classical states of light is bounded by the standard-quantum-limit (SQL) $\Delta \phi > \Delta \phi_{\text{SQL}} =1/\sqrt{M}$, where $M$ is the number of photons detected. The SQL is referenced to the total power detected and in the technical literature it is commonly called shot-noise-limit. Since, on the one hand, the enclosed area is limited by practicality and transmission loss\cite{wang:2012}, and on the other hand, the power circulating in the interferometer cannot be increased without inducing detrimental nonlinear optical effects\cite{lefevre:2014} or even damage to the system, the possibility of increasing the per-photon sensitivity of the FOG is both of fundamental and economic interest.

Quantum metrology provides a route to improve the precision of measurement to levels which would be impossible with classical resources alone \cite{dowling:2008}. Quantum interferometry thereby pursues the approach of using non-classical states of light in order to measure optical phases with a higher precision per photon. The canonical example of such states are path-entangled NOON-states, for which the resulting measurement advantage is based on the collective behavior of $N$ photons. That is, all $N$ photons are in an equal superposition of being in either one of the two modes of an interferometer, resulting in a shortened de-Broglie wavelength $\lambda/N$, where $\lambda$ denotes the physical wavelength of the individual photons\cite{jacobson:1995}. This leads to an increase of the interferometric fringe pattern frequency by a factor of $N$ (super-resolution) without changing the physical wavelength of the photons, allowing the latter to be chosen for optimized transmission through optical single-mode-fibers (SMF). 

Thus, using NOON-states, the relative phase imparted on the interferometric modes can be determined with the same precision as if consecutive photons with $N$-fold energy ($N$ times shorter wavelength), or $N$ times more photons were used. The resulting measurement advantage is called super-sensitivity, meaning that the achieved precision is better than the SQL which is a strict limit when using consecutive single photons, e.g. in coherent or thermal states with an average photon number of $M$. Using $M/N$ NOON-states one obtains a precision of $\Delta \phi_{\text{NOON}} =1/\sqrt{N M}<\Delta \phi_{\text{SQL}}$, although the same number of photons/same energy is detected. Note that the fundamental limit in quantum mechanics is the Heisenberg limit $\Delta \phi_{\text{HL}} =1/M$, which retains its validity for arbitrary photon states.

While the applicability of these non-classical states in metrology has already been widely demonstrated in other interferometric settings \cite{edamatsu:2002,Steuernagel:2002, walther:2004,Boto:2000, Giovannetti:2004}, the measurement advantage, or quantum enhancement, has not yet been used to measure a phase shift imparted by accelerated or rotational motion. In this work, we demonstrate an entanglement-enhanced phase sensitivity in a fiber optic gyroscope\cite{kolkiran:2007}. The gyroscope is based on a compact source of entangled photon pairs that was mounted together with a fiber coil on a rotating platform (see Figure \ref{fig:Mixer}). We investigated the interference signal of the counter-propagating modes in the FOG at different rotational speeds and use the canonical example of a two-photon N00N state ($N=2$) to provide phase supersensitivity beyond the standard quantum limit (SQL) in a per detected photon regime. Our work now demonstrates the experimental accessibility of entangled photon states in Sagnac interferometry and thus represents an important step towards reaching quantum-enhanced sensitivity of inertial navigation systems. 


\section*{Results}
\begin{figure*}[!htb]
\begin{figfont}
    \minipage{0.35\textwidth}
        \def\svgwidth{\linewidth}
        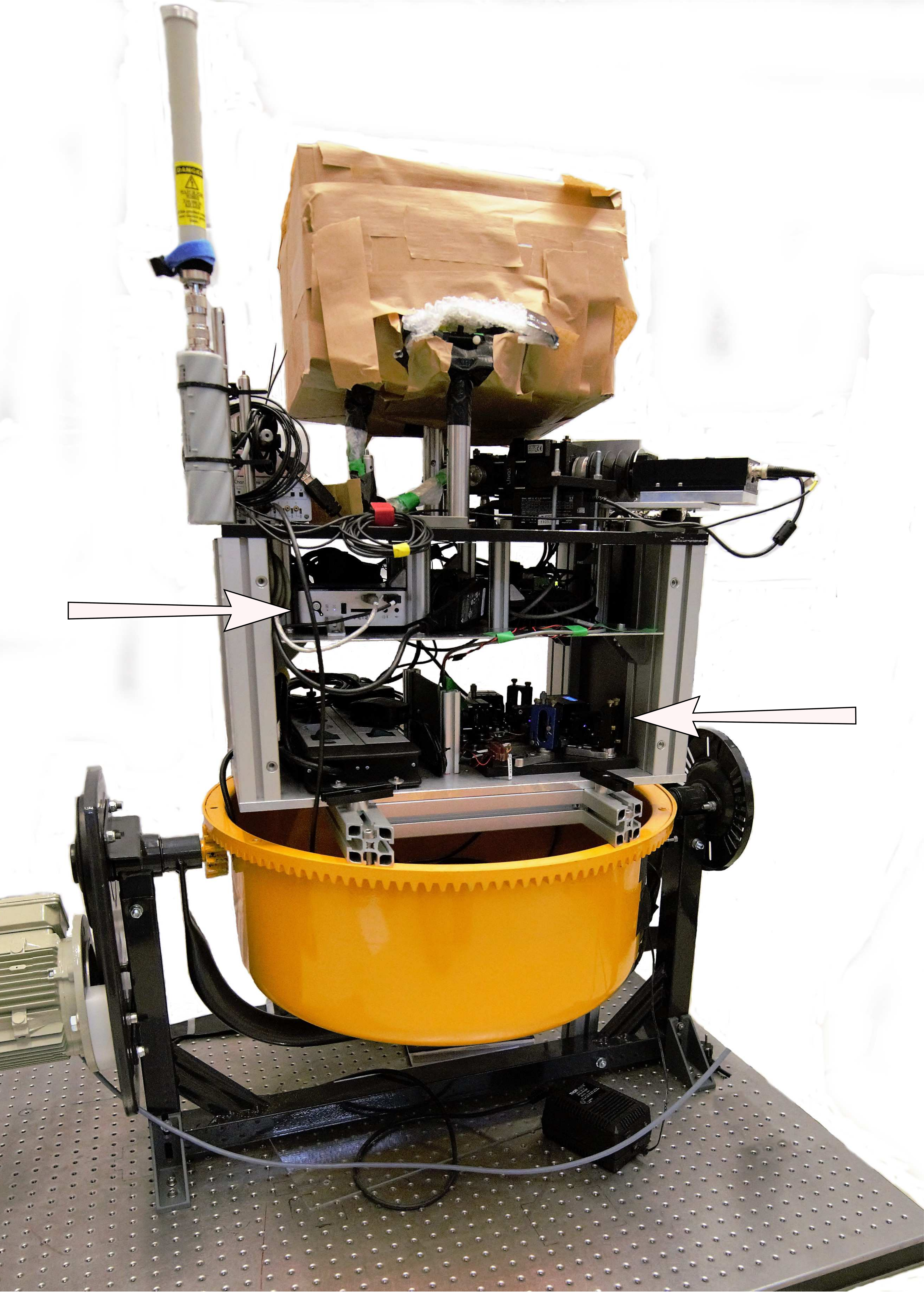
        \caption{\label{fig:Mixer}\textbf{Experimental arrangement:} A rigid three-level crate containing a photon source, a fiber based Sagnac interferometer and electronic equipment was mounted on a modified cement mixer. A remotely controlled personal computer (PC) was used for data storage and to control the experiment. The optical fiber coil of the Sagnac interferometer was passively stabilized against temperature drifts with a cardboard box and bubble wrap.}
    \endminipage\hfill
    \minipage{0.6\textwidth}
        \def\svgwidth{\linewidth}
        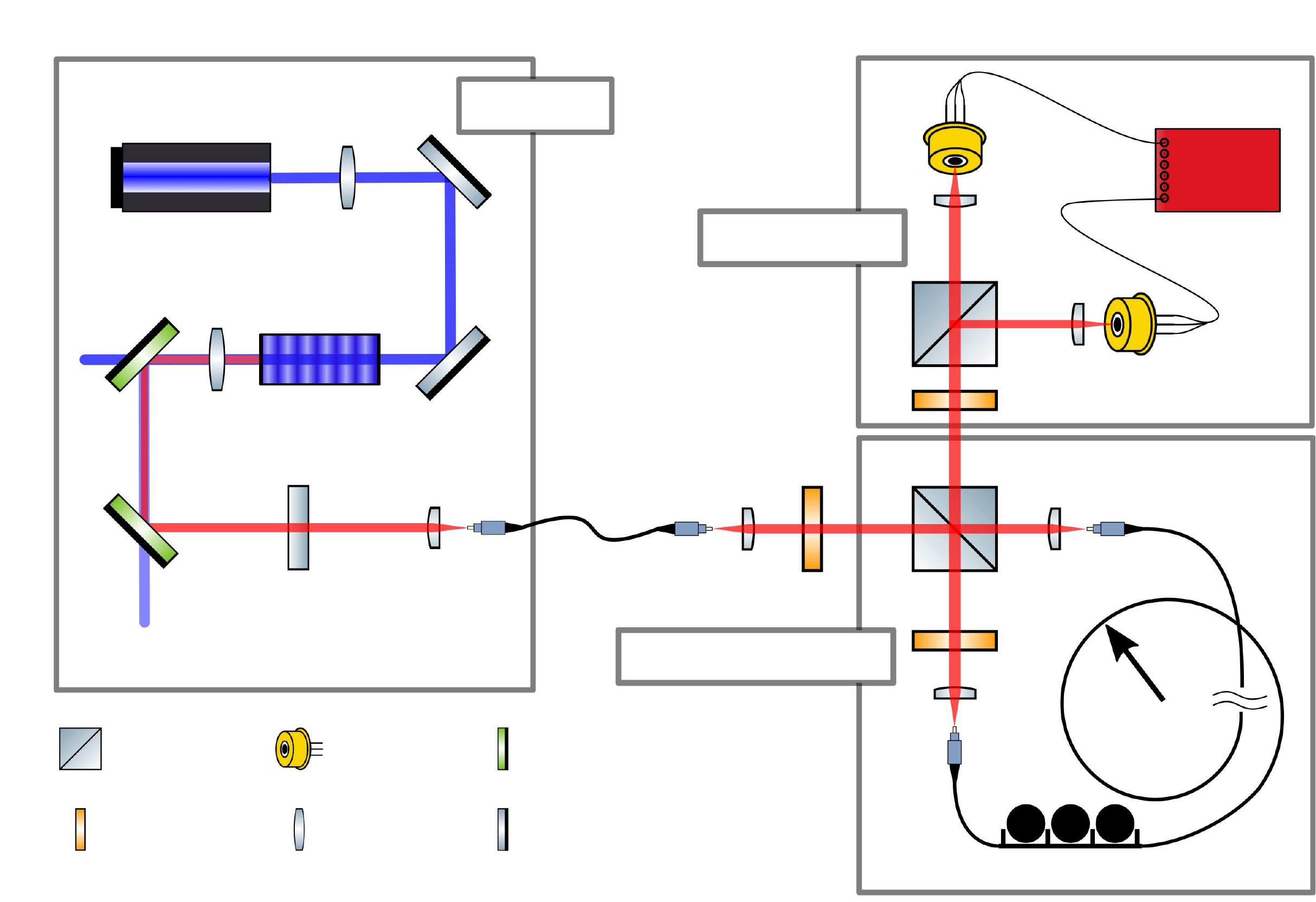
        \caption{\label{fig:optics}\textbf{Optical setup:} Light from the laser diode (405nm) produces signal and idler photons at a wavelength of $\lambda=810$ nm via spontaneous parametric down conversion (SPDC). A neodymium-doped yttrium orthovanadate (Nd:YVO4) and a polarization maintaining single-mode fiber (PMF) are used to compensate temporal walk off between the photons of a pair. A polarizing beam splitter (PBS) with ports a,b,c and d converts the polarization modes to spatial modes, which are coupled to the different ends of a single mode fiber (SMF) coil. The photons in port-d are measured in the DA-basis. Two silicon-avalanche photo diodes (SiAPD) generate electrical pulses which are recorded in time with a time tagging module (TTM).}
    \endminipage\hfill
\end{figfont}
\end{figure*}

A rigid three-level crate containing the optical setup and battery-powered electronic control equipment was mounted on a modified cement mixer. A remotely controlled computer on board was used to store all relevant data and to control the experiment. By controlling the motor by means of a variable-frequency drive, we were able to adjust the rotational speed of the Sagnac interferometer. An additional commercial gyroscope (Dytran, VibraScout 6D, 5346A2) was mounted at the cement mixer to log the rotational velocity. The optical setup (see Figure \ref{fig:optics}) generates, detects, and analyzes photon pairs and has proven its stability in previous experiments\cite{Fink:2017}. The photon source uses a continuous-wave laser at $405$ nm to produce signal and idler photons at a wavelength of $\lambda=810$ nm via spontaneous parametric down conversion (SPDC), which are subsequently collected in a polarization maintaining single-mode optical fiber (PMF). For details see Methods section. A half-wave plate (HWP) @ $22{.}5^\circ$ transforms the two-photon state at the output of the PMF $\Ket{\Psi^{\text{PMF}}}=\Ket{1_{\text{H}},1_{\text{V}}}$ to a NOON state in polarization modes ($1/\sqrt{2} (\Ket{2_{\text{H}},0_{\text{V}}} - (\Ket{0_{\text{H}},2_{\text{V}}})$), where both photons are either horizontal (H) or vertical (V) polarized.

Additionally, for providing reference measurements with consecutive single photons, the one-photon state $\Ket{1_{\text{H}},0_{\text{V}}}$ could be generated by inserting a horizontally oriented polarizer directly after the PMF. After the transmission through a HWP, the single photon is anti-diagonal polarized and its state can be written as a superposition of H and V polarization $\Ket{0_{\text{D}},1_{\text{A}}}=1/\sqrt{2} (\Ket{1_{\text{H}},0_{\text{V}}} - \Ket{0_{\text{H}},1_{\text{V}}})$. The polarizing beamsplitter (PBS) of the Sagnac interferometer (see Fig. \ref{fig:optics}) converts the polarization modes to spatial modes, which are both coupled to the different ends of a coiled fiber loop. Transmitted photons thus propagate in a clockwise ($\lcirclearrowright$) and reflected in a counterclockwise ($\rcirclearrowright$) direction through the fiber. The resulting $N$-photon state in the Sagnac interferometer is given by

 \begin{equation}
 \frac{1}{\sqrt{2}}\left(\Ket{N_\lcirclearrowright,0_\rcirclearrowright} - \Ket{0_\lcirclearrowright,N_\rcirclearrowright}\right),
    \label{Eq:NOON}
\end{equation}
with $N$ being either 1 or 2 in case of a measurement with or without the polarizer, respectively. After traversing the fiber loop, the photons impinge on the PBS a second time, from where they are guided to output port-d. The polarization changes in the fiber-coil where compensate by an additional HWP and an in-fiber-polarization-controller (see Figure \ref{fig:optics}). 

The rotation with an angular velocity $\Omega$ leads to a phase difference $\phi_S$ between counter propagating waves. Ignoring relativistic corrections due to the slow rotational velocity in our experiment, the Sagnac phase\cite{gift:2018} reads:

\begin{equation}
        \phi(\Omega)=c \frac{2 \pi}{\lambda} \Delta t (\Omega) \simeq \frac{4 \pi L r \Omega}{\lambda c} = S_{T} \Omega = 1.09 \Omega,
    \label{Eq:PhSag}
\end{equation}

where $c$ denotes the speed of light, $r\sim7{.}8$ cm is the radius of the fiber coil, $L\sim270{.}5$ m is the length of the fiber, and $\lambda=810$ nm is the wavelength of the down-converted photons. The final state after transmission through the Sagnac loop (in port-d of the PBS) reads: 

\begin{equation}
    \Ket{\Psi^{\text{port-d}}(N)} = \frac{1}{\sqrt{2}}\left(\Ket{N_{\text{H}},0_{\text{V}}} - \mathrm{e}^{i N \left(\phi(\Omega)+\phi_0\right)} \Ket{0_{\text{H}},N_{\text{V}}}\right).
 \label{Eq:PortD}
\end{equation}

where $\phi_0$ accounts for an initial offset caused by the birefringence of the fiber coil. The detection module consists of a HWP at $22{.}5^\circ$ followed by a PBS and two silicon-avalanche photo diodes (SiAPD), one in each output. This setup projects the state $\Ket{\Psi^{\text{port-d}}(N)}$ onto the diagonal/anti-diagonal (D/A) polarization basis. The which-way information ($\rcirclearrowright$/$\lcirclearrowright$) is inaccessible in this basis, resulting in interference between the two modes. Note that the $N$-photon rate $R_{N} \propto \cos ^2\left(N/2 \left(S\;\Omega+\phi_0\right)\right)$ shows a linear increase of the fringe oscillation frequency $\omega_N=(S N)/2$ with $N$.

\begin{figure*}
    \minipage{0.49\textwidth}
    \begin{figfont}
    \begin{subfigure}[b]{\linewidth}
            \replacenumber{\scalefigfont}
            \replacepi{\scalefigfont}
            \psfrag{T}[][][\scalefigfont]{Time [s]}
            \psfrag{NC}[][][\scalefigfont][0]{$R_{1}$ per 5 ms}
            \psfrag{O}[cl][cr][\scalefigfont][0]{$\Omega$ [rad s$^{-1}]$}
            \begin{overpic}[width=\linewidth]{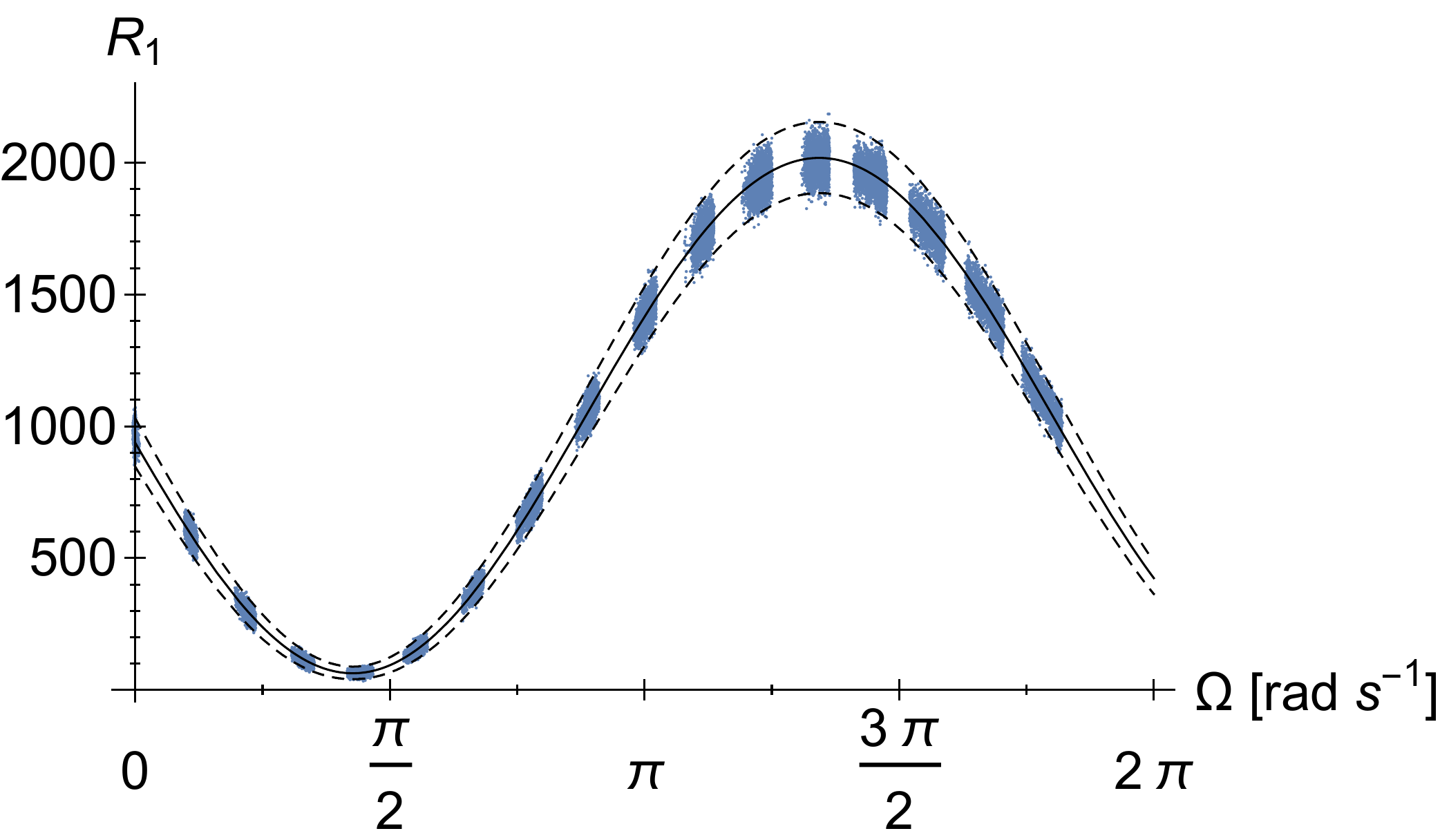}
            \end{overpic}
            \caption{\label{fig:NSi}}
    \end{subfigure}\hfill
    \end{figfont}
    \endminipage\hfill
    \minipage{0.49\textwidth}
    \begin{figfont}
        \begin{subfigure}[b]{\linewidth}
            \replacenumber{\scalefigfont}
            \replacepi{\scalefigfont}
            \psfrag{NC}[][][\scalefigfont][0]{$R_{2}$ per 20 ms}
            \psfrag{O}[cl][cr][\scalefigfont][0]{$\Omega$ [rad s$^{-1}]$}
            \begin{overpic}[width=\linewidth]{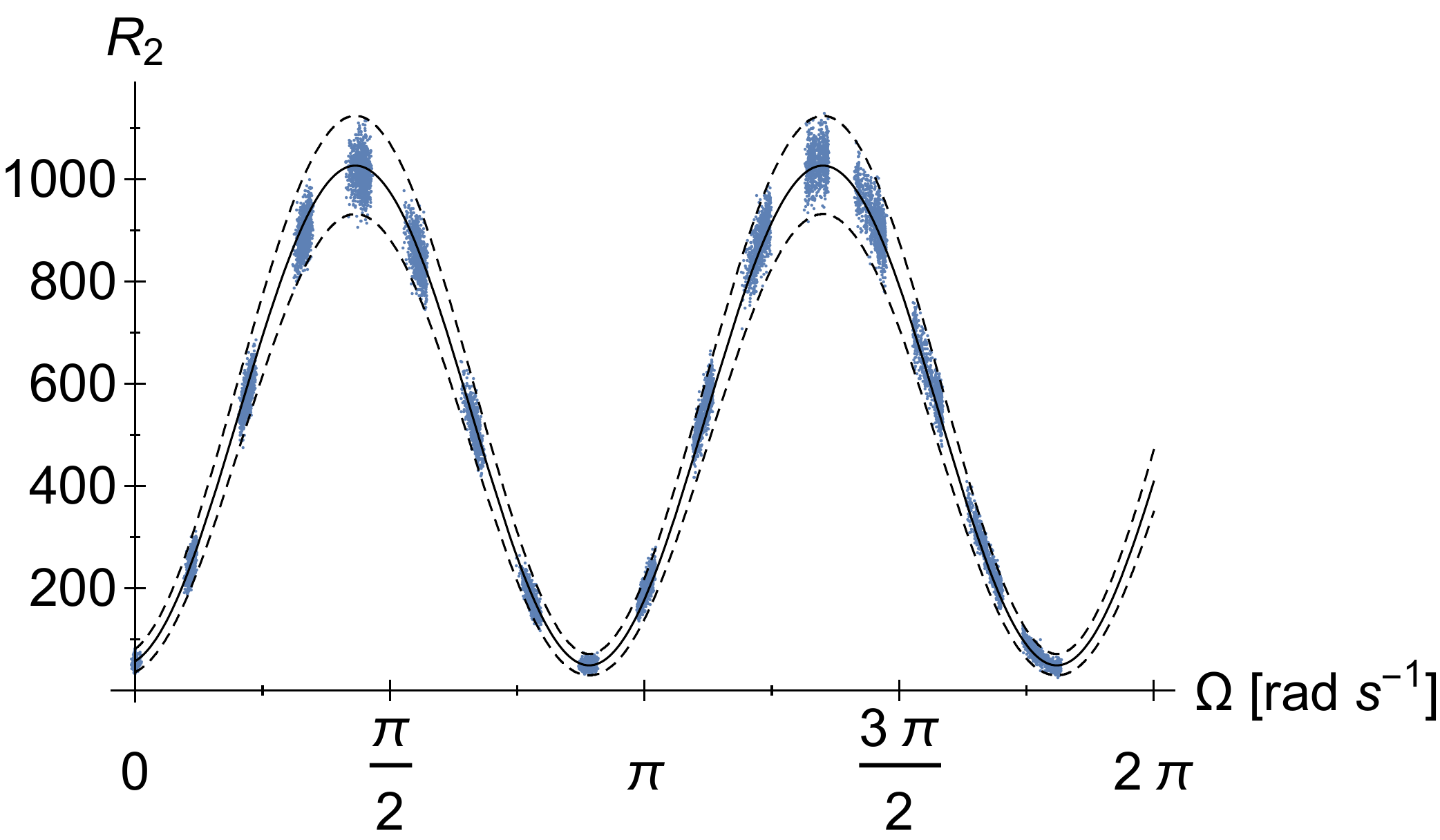} 
            \end{overpic}
            \caption{\label{fig:NCoi}}
        \end{subfigure}
    \end{figfont}
    \endminipage\hfill
\caption{\label{fig:N} \textbf{Count rate vs. angular velocity}: Rate of $N=1$/$N=2$ photon state detection events (\subref{fig:NSi}/\subref{fig:NCoi}). The blue dots indicate the experimentally determined number of single-photon and two-photon counts per second. Each of the $64{,}688$/$16{,}204$ points was measured over $\tau_1=5$/$\tau_2=20$ ms. The solid black line is a fit function of the form of equation (\ref{Eq:fit}). The dashed lines indicate the 99\% confidence interval around the fitted function if only Poissonian error of the predicted count rate is assumed. Those lines contain $97{.}25$\%/$95{.}88$\% of the measured data.}
\end{figure*}

Two sequential measurement runs were performed using a one-photon state ($N=1$) and a two-photon NOON-state ($N=2$), with and without the polarizer, respectively. For each run, the angular velocity was increased step wise from $\Omega=0$ to $\Omega=5{.}6$ rad per second, resulting in a full $2\pi$ phase shift according to equation (\ref{Eq:PortD}). For each setting of $\Omega$ we accumulated data for a total of $\simeq 19$ seconds and evaluated the one-photon/two-photon count rates $R_{1}$/$R_{2}$ with integration times $\tau_1=5$ ms / $\tau_2=20$ ms, leading to 3800/950 data-points per angular velocity setting. Note that using different integration times for the $N=1$ and $N=2$ measurements accounts for the different rates of detected one-photon and two-photon states and thus results in roughly the same number of detected photons per data point for both measurements. Figure \ref{fig:NSi} shows the one-photon count rate $R_{1}$ of $\text{SiAPD}_2$ obtained in the first measurement run, whereas Figure \ref{fig:NCoi} shows the two-photon coincidence count rate $R_{2}$ between $\text{SiAPD}_1$ and $\text{SiAPD}_2$ of the second measurement run. Comparison of the two graphs clearly shows the increased fringe oscillation frequency of the coincidence count rate $R_2$. The data in Figure \ref{fig:N} are fitted with a function of the form of equation (\ref{Eq:fit}) and using the fitting parameters presented in Table \ref{tab:fitPar}. The error estimation of the fit parameters was determined via bootstrapping and confirmed by Monte Carlo simulation.

\begin{equation}
    R_{N}=  M/N  \cos ^2\left(\frac{N}{2} \left(S \Omega+\phi_0\right)\right) + B
    \label{Eq:fit}
\end{equation}

\begin{table}[h]
\begin{center}
    \begin{tabular}{c|cccc}
     & $M$ & $B$ & $S$ & $\phi_0$ \\
    \hline
    $N=1$ & $1955$ & $63$ & $1.091(8)$ & $1.676(7)$ \\
    $N=2$ & $1956$ & $49$ & $1.089(0)$ & $1.662(9)$ \\
    \end{tabular}
  \caption{\label{tab:fitPar}Fit parameters of equation (\ref{Eq:fit}) as plotted in Figure \ref{fig:N}.}
\end{center}
\end{table}

The phase shift caused by the fiber coil $\phi_0$ was set to $\sim \pi/2$ using the fiber-polarization-controller. The background $B$ results mainly from uncorrelated residual background counts but also from imperfect overlap of the counter propagating modes. Note that the amplitude in Figure \ref{fig:NSi} is about twice as large as in Figure \ref{fig:NCoi}, but  $M\sim 2000$ is approximately the same for both measurement runs. Here, $M$ denotes the total number of photons detected per integration time $\tau$, in the respective photon state. The fiber coil is not perfectly circularly wound and has extra loops for the polarization controller and the PBS connections, therefore equation (\ref{Eq:PhSag}) is a rough estimation of the expected scale factor $S_{T}$, which agrees nicely with the fitted parameter $S$.

\begin{figure}[h]
    \centering
    \begin{figfont}
            \replacenumber{\scalefigfont}
            \psfrag{NC}[][][\scalefigfont][180]{$R_{2}$ per 20 ms}
            \psfrag{O}[][][\scalefigfont][180]{$\Omega$ [rad s$^{-1}]$}
            \includegraphics[width=0.5\textwidth]{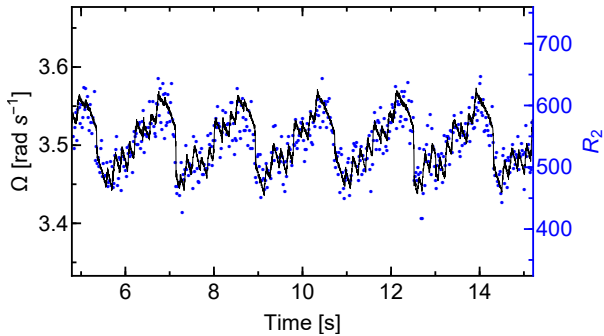}
    \end{figfont}
\caption{\label{fig:CoiTime}  \textbf{Two-photon count rate $R_{2}$ and angular velocity vs. time}: Variation in time of the angular velocity $\Omega$ as measured with the commercial gyro-sensor (black solid line). The blue dots indicate the two-photon count rate $R_{2}$, with integration time of $20$ ms.}
\end{figure}

The vertical distribution of the individual measurements in Figure \ref{fig:N} with respect to the fit function is mainly caused by statistical count-rate fluctuations. The horizontal distribution of the measurements stems from variations in the rotational velocity of the modified cement mixer (due to an imbalanced ball bearing mechanism). However, knowledge on the timing of individual detection events, we were able to directly resolve the variations in rotation period. The correlation of the coincidence count rate and the rotational velocity as measured by the commercial gyro sensor is shown in Figure \ref{fig:CoiTime}.

\begin{figure*}
    \begin{figfont}
    \begin{subfigure}[b]{1\textwidth}
            \replacenumber{\scalefigfont}
            \replacepi{\scalefigfont}
            \psfrag{T}[][][\scalefigfont]{Time [s]}
            \psfrag{NC}[][][\scalefigfont][0]{$R_{1}$ per 5 ms}
            \psfrag{O}[cl][cr][\scalefigfont][0]{$\Omega$ [rad s$^{-1}]$}
            \begin{overpic}[width=0.9\textwidth]{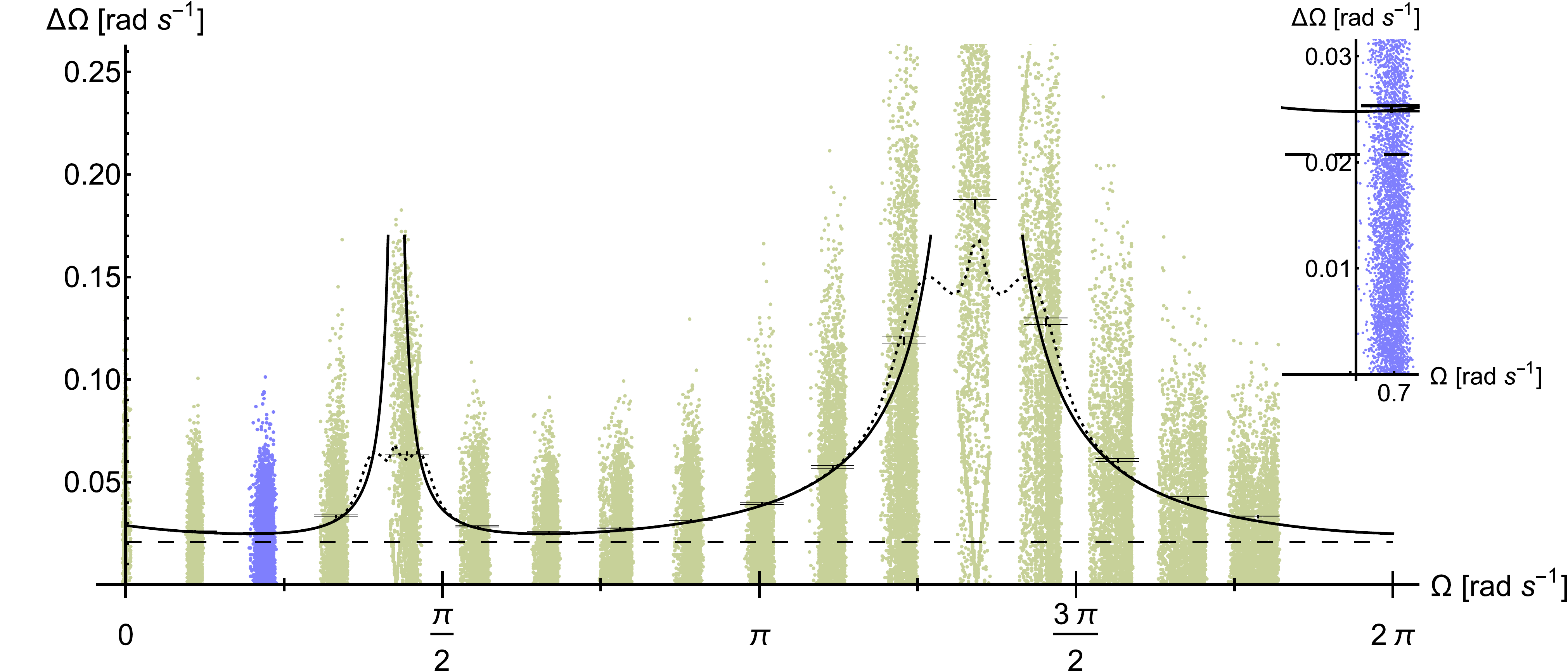}
            \end{overpic}
            \caption{\label{fig:ESi}}
    \end{subfigure}
    \end{figfont}
        \vspace{.1in}
    \begin{figfont}
        \begin{subfigure}[b]{1\textwidth}
            \replacenumber{\scalefigfont}
            \replacepi{\scalefigfont}
            \psfrag{NC}[][][\scalefigfont][0]{$R_{2}$ per 20 ms}
            \psfrag{O}[cl][cr][\scalefigfont][0]{$\Omega$ [rad s$^{-1}]$}
            \begin{overpic}[width=0.9\textwidth]{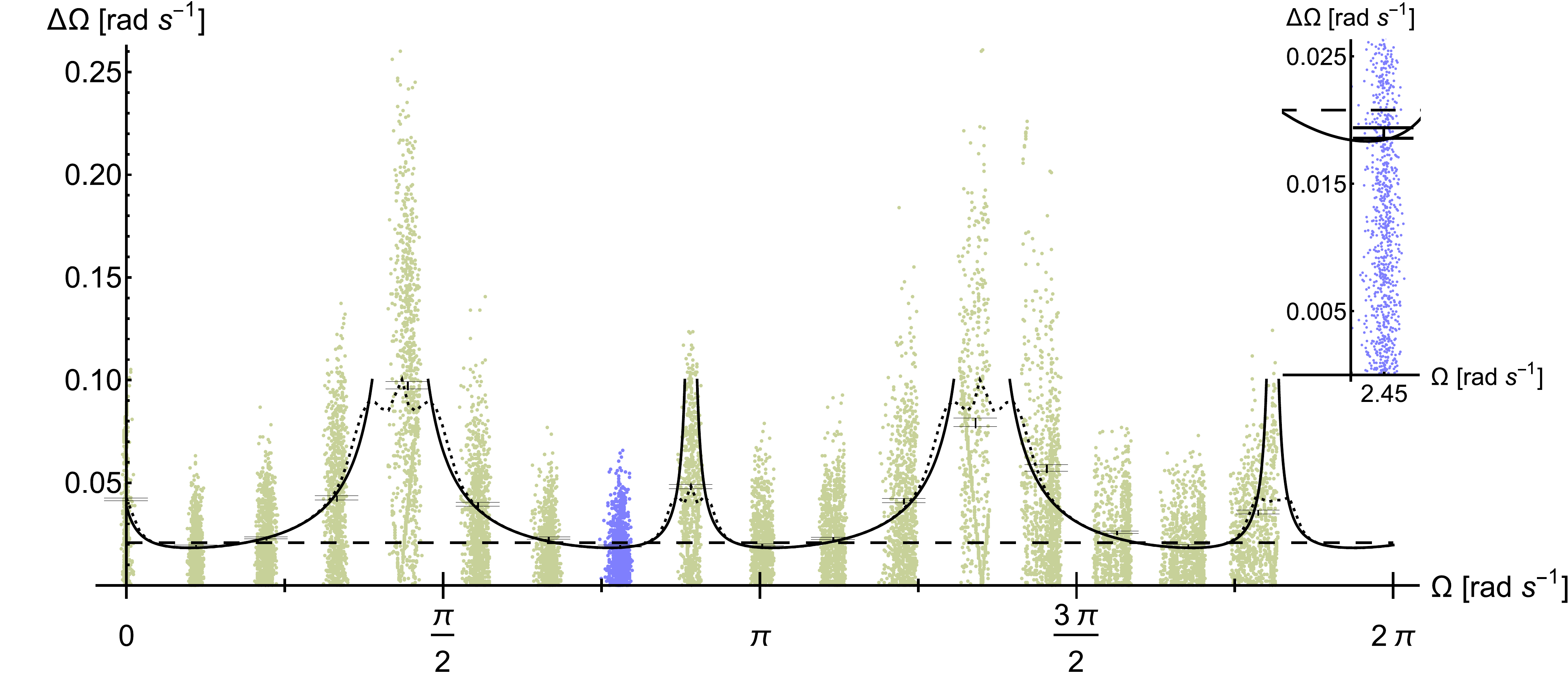} 
            \end{overpic}
            \caption{\label{fig:ECoi}}
        \end{subfigure}
    \end{figfont}
    \caption{\label{fig:GyroSQL} 
        \textbf{One-photon vs two-photon measurement precision}.
        The precision of the rotational velocity, measured with the one-/two-photon state is shown in (a)/(b) respectively.
        Absolute value of the Deviation $|\Omega^E(R_N^i)- \Omega^i|$ (green dots). Sample standard deviation of $\left[\Omega^E(R_N^i)- \Omega^i\right]$ for each measurement block $\Delta \Omega$ with estimated errors (black error bars). SQL of the measurement system $\Delta \Omega^\text{SQL}$ (dashed line). Uncertainty estimated via fit function $\Delta \Omega^E (R_N(\Omega))$ (solide black line) and numerically estimated (dotted black line). The measurements with the smallest standard deviation $\Delta \Omega_\text{min}$ is colored blue. These data are shown enlarged in the upper right corner.
    }
\end{figure*}

The uncertainty in the measured velocity attributed to Poisson noise of $R_N$ can be estimated via propagation of uncertainty $\Delta \Omega^E (R_N) = |\partial \Omega^{E}(R_N)/\partial R_N| \sqrt{R_N }$, where $\Omega^E (R_N)$ is the rotational velocity estimated via the inverse fit function (equation (\ref{Eq:fit})). The uncertainty calculated this way is plotted along with a numerically calculated uncertainty in Figure \ref{fig:GyroSQL}, black solid and dotted line, respectively. Usually, such gyroscopes are operated at the point of best precision, which is called bias-point. The expected precision at the bias-point $\Delta \Omega^E_\text{Bias}$ is numerically compared to $\Delta \Omega^\text{SQL} = 1/(S \sqrt{M})$ and $\Delta \Omega^\text{HL} = 1/(S M)$ in Table \ref{tab:GyroSQL}. 

With the inverse fit function, a rotational velocity $\Omega^E(R_N^i)$ can be assigned to each measured count rate (blue points in Figure \ref{fig:NSi} and \ref{fig:NCoi}). Those values deviate from the reference $\Omega$, as measured with the commercial gyro-sensor. The absolute value of those deviations are plotted in Figure \ref{fig:GyroSQL} (green points), together with the sample standard deviation (black error bars) of each block of measurements (measurements with similar rotational velocity). The block with the best precision of each measurement run is coloured blue and the respective sample standard deviation $\Delta \Omega_\text{min}$ can be found in Table \ref{tab:GyroSQL}.

\begin{table}[h]
\begin{center}
    \begin{tabular}{c|cccccc}
     & $\Delta \Omega^\text{SQL} $& &$\Delta \Omega_\text{min} $ & $\Delta \Omega^E_\text{Bias}$ & & $\Delta \Omega^\text{HL} $ \\
    \hline
    $N=1$ & \multirow{2}{*}{$ 0{.}0207$ }& $\rlap{\kern.3em$ / $}>$ & $0.025(0)$ & $0.0248$ & $>$ & \multirow{2}{*}{$469 \times 10^{-6}$ } \\
    $N=2$ & & $>$ & $0.018(9)$ & $0.0183$ & $>$ \\
    \end{tabular}
  \caption{\label{tab:GyroSQL} Standard quantum limit and Heisenberg limit of the measurement system $\Delta \Omega^\text{SQL}$ and $\Delta \Omega^\text{HL}$, respectively. Best precision of the respective measurement run $\Delta \Omega_\text{min}$ (blue colored data in Figure \ref{fig:GyroSQL}). Precision at the bias point $\Delta \Omega^E_\text{Bias}$, estimated via fit function. All values are given in rad per sec. }
\end{center}
\end{table}

\section*{Discussion}

In summary, we have demonstrated the effect of super-resolution in a FOG using two-photon NOON-states. The increased fringe oscillation frequency of the two-photon count rate $R_2$ with respect to the one-photon count rate $R_1$, is shown in Figure \ref{fig:N}. From the fit parameters of Table \ref{tab:GyroSQL}, a fringe oscillation frequency ratio of $\omega_{N=2}/\omega_{N=1} = S_{N=2}/(\tfrac{1}{2}S_{N=1}) = 1.995 (\pm 0.4 \times 10^{-3})$ can be found. This value deviates from the theoretically expected value 2. The deviation is on the order of 10 standard deviation and therefore not fully covered by the specified error, which may be due to systematic errors such as drift of the birefringence of the fiber coil or scale factor $S$ variations. Nevertheless, the inconsistency is rather small, given the poor thermal and mechanical stabilization of the fiber. However, the ratio is significantly greater than 1, by means of 2487 standard deviations, which bears witness of super-resolution.

Furthermore we have investigated the precision of the NOON-state measurements close to the bias-point $\Delta \Omega_\text{min} $ (see Table \ref{tab:GyroSQL}). We found that, mainly due to background, it is slightly worse than the theoretical predicted limit $\Delta \Omega^\text{SQL}/\sqrt{2} = 0.0146$ rad $s^{-1}$. Nevertheless, the measured precision of the two-photon state performs better than the one-photon state, and is better than $\Delta \Omega^\text{SQL}$. Such sub-SQL precision can never be achieved with coherent states of light, and demonstrates super-sensitivity.

In conclusion our result suggest a possible improvement of the accuracy of FOG using non-classical states of light. In particular, when using NOON-states with large $N$, one can significantly reduce the de-Broglie wavelength of the photon state, without altering the wavelengths of the actual photons. Hence the de-Broglie wavelength can potentially be shifted outside of the transmission window of the FOG. At this point we should stress, however, that the presented technology is not yet competitive with a classical FOG. Laser-driven FOG\cite{chamoun:2014} use an optical power of approx. $20\mu$W, which corresponds to a rate of $156 \times 10^{12}$ photons per second (at $\lambda = 1550$ nm). In contrast, the detected photon rate of the NOON state is $100 \times 10^3$ in our experiment. This relatively low photon rate was limited by the detectors used, whose efficiency decreases with increasing count rate. 

Hence, while our sensor is not yet competitive with commercial gyroscopes, we believe that our work can be considered an important first step towards reaching the ultimate sensitivity limits in Sagnac interferometry. With experiments and applications becoming increasingly demanding --- so does the severity of the limitation imposed by Shot noise. Moreover, since power circulating in the interferometer cannot be increased arbitrarily, due to detrimental power-dependent effects such as to coherent back-scattering\cite{chamoun:2014} or nonlinear Kerr effects\cite{lefevre:2014}, methods and techniques from quantum metrology will play a significant role in reaching the ultimate sensitivity limits of FOG and enable evermore demanding applications in fundamental science and technology. Wit the speed of ongoing developments in advancing detector technology and increasingly brighter photon sources, a technical application of such a system may become feasible in the foreseeable future.  We hope that our work will inspire further research in this direction.

\begin{methods}

\subsection{Photon source}
A continuous-wave laser at $405$ nm is focused into a periodically poled Potassium Titanyl Phosphate (ppKTP) crystal. The photons from the pump laser are converted within the crystal via spontaneous parametric down conversion (SPDC) into pairs of signal and idler photons with horizontal (H) and vertical (V) polarization, respectively. In order to guarantee wavelength-degenerate quasi-phase matching at 810 nm, the temperature of the crystal is stabilized at $37{.}375^{\circ}\text{C}\pm 0{.}01^{\circ}\text{C}$. The wavelength-dependent splitting is realized with to dichoric mirrors. While the signal and idler photons are coupled into one PMF, the rest of the pump light is guided to a CMOS camera for the purpose of stability control. The birefringence of the ppKTP crystal leads to a longitudinal walk-off between the down converted photons. In order to create a NOON-State with two indistinguishable photons, one has to compensate this polarisation dependent time delay with further birefringent components of the right length. For this purpose, we have used an additional neodymium-doped yttrium orthovanadate (Nd:YVO4) crystal, considering the birefringence of the PMF.
This source produces a measured photon rate of $N_S=950{,}000$ sec${^{-1}}$ at $810$ nm with a pump power of $27{.}5$ mW. Because of photon losses within the system, only a fraction of the photons leaves the PMF as a pair. We measure a photon pair rate of $191{,}000$ sec${^-1}$. Both rates are measured with a photon detection efficiency of $P_\text{SiAPD}=0{.}64$ (data sheet of the manufacturer).

\end{methods}

\bibliography{sample}

\begin{addendum}
 \item[Acknowledgements] We would like to thank Roland Blach for preparing and testing the concrete mixer. Financial support from the Austrian Research Promotion Agency (FFG contract 854022, 866025 and 844360) as well as the Austrian Academy of Sciences is gratefully acknowledged. J.P.D. would like to acknowledge support from the Air Force Office of Scientific Research, the Army Research Office, the Chinese academy of science, the Defense Advanced Research Projects Agency, the National Science Foundation, and the Northrop Grumman Corporation.
 \item[Correspondence] Correspondence and requests for materials should be addressed to \newline M.F.~(email: Matthias.Fink@oeaw.ac.at) and \newline 
 R.U.~(email: Rupert.Ursin@oeaw.ac.at).
\end{addendum}

\end{document}